\documentclass[twocolumn]{physlett}
\usepackage{graphics}
\usepackage{times}
\graphicspath{{.}{./pictures/}}
\providecommand{\Ed}{\ensuremath{E_\mathrm{d}}}
\providecommand{\Ld}{\ensuremath{L_\mathrm{d}}}
\begin{document}
\begin{frontmatter}

\title{Coherent $\pi^0$ threshold production from 
  the deuteron at $Q^2 = 0.1\,\mathrm{GeV^2/c^2}$}

\author[kph]{I.~Ewald\thanksref{thesis}}, \author[kph]{P.~Bartsch},
\author[kph]{D.~Baumann},                 \author[phy]{J.~Bermuth},            
\author[mit]{A.\,M.~Bernstein},           \author[jss]{K.~Bohinc},
\author[kph]{R.~B\"ohm},                  \author[kph]{N.~Clawiter},
\author[kph]{S.~Derber},                  \author[kph]{M.~Ding},
\author[kph]{M.\,O.~Distler},             \author[kph]{J.~Friedrich},
\author[kph]{J.\,M.~Friedrich},           \author[kph]{M.~Kahrau},
\author[tud]{M.~Kohl},                    \author[mel]{A.~Kozlov},
\author[kph]{K.\,W.~Krygier},             \author[kph]{A.~Liesenfeld}, 
\author[kph]{H.~Merkel\thanksref{mail}},  \author[kph]{P.~Merle},
\author[kph]{U.~M\"uller},                \author[kph]{R.~Neuhausen},
\author[mit]{M.~Pavan},                   \author[kph]{Th.~Pospischil},
\author[jss]{M.~Potokar},                 \author[phy]{D.~Rohe},
\author[kph]{G.~Rosner},                  \author[kph]{H.~Schmieden},
\author[jss]{S.~\v{S}irca},               \author[kph]{A.~Wagner},
\author[kph]{Th.~Walcher},           \and \author[kph]{M.~Weis}

\address[kph]{Institut f\"ur Kernphysik,
  Johannes Gutenberg-Universit\"at Mainz, D-55099~Mainz, Germany}
\address[phy]{Institut f\"ur Physik,
  Johannes Gutenberg-Universit\"at Mainz, D-55099~Mainz, Germany}
\address[mit]{Laboratory for Nuclear Science, 
  Massachusetts Institute of Technology, Cambridge, MA~02139, U.S.A.}
\address[jss]{Jo\v{z}ef Stefan Institute, SI-1001~Ljubljana, Slovenia}
\address[tud]{Institut f\"ur Kernphysik, 
  TU Darmstadt, D-64289 Darmstadt, Germany}
\address[mel]{School of Physics, The University of Melbourne, 
  Victoria 3010, Australia}
\thanks[thesis]{Comprises parts of the Doctorate thesis of I.~Ewald}
\thanks[mail]{Corresponding author, 
  E-mail:~Merkel@KPh.Uni-Mainz.de,
  Phone:~+49\,(61\,31)\,39\,-\,25812, Fax:~-\,22964}
\date{\today}

\begin{abstract}
  First data on coherent threshold $\pi^0$ electroproduction from the
  deuteron taken by the A1 Collaboration at the Mainz Microtron MAMI are
  presented. At a four-momentum transfer of $q^2=-0.1\,\mathrm{GeV}^2/c^2$ the
  full solid angle was covered up to a center-of-mass energy of 4\,MeV above
  threshold. By means of a Rosenbluth separation the longitudinal threshold
  $s$ wave multipole and an upper limit for the transverse threshold $s$ wave
  multipole could be extracted and compared to predictions of Heavy Baryon
  Chiral Perturbation Theory.
\end{abstract}
\begin{keyword}
  Pion electroproduction; Threshold production; Deuteron
  \PACS{25.30.Rw; 13.60.Le; 12.39.Fe}
\end{keyword}
\runauthor{I.~Ewald \etal}
\end{frontmatter}
%%%%%%%%%%%%%%%%%%%%%%%%%%%%%%%%%%%%%%%%%%%%%%%%%%%%%%%%%%%%%%%%%%%%%%%%%%%
\section{Introduction}

The electroproduction of neutral pions is one of the significant testing
grounds of Chiral Perturbation Theory (ChPTh, see \cite{chpt94,chpt97} for an
overview of the field).  The data of the production from the proton
\cite{nikhef,distler} are consistent with calculations of Heavy Baryon Chiral
Perturbation Theory \cite{BKMe96} and are in reasonable agreement with this
approach for the free nucleon case. For this comparison still six low energy
constants, introduced into ChPTh by renormalizing counter terms, had to be
adjusted to the data. However, for the free neutron a strong prediction
without further freedom can be deduced from the proton production amplitude.

Unfortunately, no free neutron target exists and one has to cope as normal
with the problem of a target bound in a complex nucleus. Therefore, a model is
needed to separate the elementary production amplitudes and the nuclear
effects introducing theoretical uncertainties. The most evident choice for
investigating the neutron bound in a nucleus would be the quasi-free
reaction. The momenta of the recoiling nucleons and the initial Fermi momenta
are, however, of the same order and, therefore, the dynamics of the full
reaction are not well under control. On the other hand, assuming that the
proton production amplitude is known and that a reasonable model for the
simplest nucleus is at hand today, the coherent production from the deuteron
can be used.
In this case, the initial and final Fermi motion can be included
in the model description of the deuteron via structure form factors.

For the photoproduction case, a first coherent threshold measurement of
$d(\gamma,\pi^0)d$ at SAL \cite{Berg98} extracted the transverse threshold $s$
wave multipole $|\Ed|$ in good agreement with the predictions of a fourth
order calculation in the framework of ChPTh \cite{BBLM97}. In photoproduction
the reaction is identified by the detection of the decay photons of the
neutral pion, and the small incoherent break up channel has to be subtracted
by model dependent assumptions. By contrast in an electroproduction
experiment the recoil deuteron has to be detected due to large background of
hard and soft photons, thus there is no incoherent contribution. The low
kinetic energy of the heavy recoil deuteron
($25\,\mathrm{MeV}<T_\mathrm{d}<35\,\mathrm{MeV}$) is the biggest challenge of
such an experiment, the detection efficiencies and multiple scattering effects
have to be measured at each state of the experiment.

On the other hand, the detection of the deuteron with a high resolution
spectrometer allows for a good invariant mass resolution close to
threshold. The focusing due to the Lorentz boost permits to cover the full
solid angle in the center-of-mass system up to 4\,MeV above threshold if one
uses the large solid angle magnetic spectrometer A of of the A1 Collaboration
with $\Delta\Omega=21\,\mathrm{msr}$.

\section{Kinematics}
\begin{figure}
  \center \resizebox{7.5cm}{!}{\includegraphics{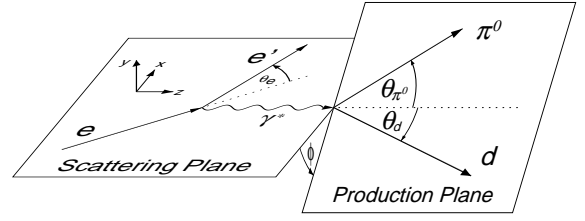}}
  \caption{Definition of angles.}
  \label{fig:kinematics}
\end{figure}

In Born approximation the virtual photon is defined by the four-vectors
of incident electron $e=(E,\vec{k})$ and scattered electron
$e'=(E',\vec{k}')$ as $q = e-e' = (\omega, \vec{q})$ (see
figure~\ref{fig:kinematics}).
Denoting the variables in the photon-deuteron center-of-mass frame with an 
asterisk the 
cross section can be written as
\begin{eqnarray}
\frac{d\sigma}{dE'd\Omega'd\Omega_\pi^*}
  &=& \Gamma \frac{d\sigma}{d\Omega_\pi^*}
\end{eqnarray}
with the virtual photon flux
\begin{eqnarray}
\Gamma &=& \frac{\alpha}{2\pi^2} 
         \frac{E'}{E} 
         \frac{k_\gamma}{-q^2}
         \frac{1}{1-\epsilon}
\end{eqnarray}
and the equivalent photon energy $k_\gamma = (W^2-m_{d}^{2})/2m_{d}$. The
transverse and longitudinal degree of polarization of the virtual photon is
given by
\begin{eqnarray}
  \epsilon            & = & \left(1-\frac{2(\omega^2 - q^2)}{q^2}
    \tan^2\frac{\theta_{e}}{2} \right)^{-1}\\
  \epsilon_\mathrm{L} & = & \frac{-q^2}{\omega^{*2}}\epsilon.
\end{eqnarray}
For the coherent production of pseudo scalar mesons from a spin 1 target
the unpolarized differential cross section can be separated similarly to
the production off a proton into four structure functions (see
e.g. \cite{eberts,aren}): 
\begin{eqnarray}
  \frac{d\sigma(\theta_\pi^*,\phi_\pi^*)}{d\Omega^*_\pi}
  &=& f_{\mathrm T}(\theta_\pi^*) +
  \epsilon_{\mathrm L} f_{\mathrm L}(\theta_\pi^*) \nonumber \\
  &+& \sqrt{2\epsilon_{\mathrm L}(1-\epsilon)}
  \cdot f_\mathrm{TL}(\theta_\pi^*)\cdot \cos\phi_\pi^* \nonumber \\
  &+& \epsilon \cdot f_\mathrm{TT}(\theta_\pi^*) \cdot \cos 2\phi_\pi^*.
\end{eqnarray}
In principle, the extraction of the transverse-longitudinal $f_\mathrm{TL}$
and transverse-transverse interference structure functions $f_\mathrm{TT}$ is
possible by a measurement of the $\phi_\pi^*$ dependence of the differential
cross section. In the presented experiment, however, the angular resolution of
the detection of the low energy deuterons was not sufficient to allow an
extraction of these small structure functions.

The choice of the four-momentum transfer of $q^2\,=\,-0.1\,\mathrm{GeV^2/c^2}$
as a lower limit was dictated by the detection efficiency of the deuteron as
will be discussed in the next section.  For this four-momentum transfer we
chose three different values for the virtual photon polarization $\epsilon$
with the largest possible spread for a Rosenbluth
separation. Table~\ref{tab:kinematics} summarizes the kinematical settings.

\section{Experimental Setup and Analysis}

\subsection{Particle Detection and Efficiencies}

The experiment was performed at the three spectrometer setup of the A1
Collaboration at the MAMI accelerator (see ref.~\cite{dreispek} for a detailed
description of the setup). For the electron detection spectrometer B, 
a clamshell dipole spectrometer with an angular acceptance of 5.6\,msr at a
momentum resolution of $\Delta p/p = 10^{-4}$ was used. For the deuteron
detection spectrometer A with a large solid angle of 21\,msr was chosen.

\begin{table}
  \caption{Kinematical settings. The four-momentum transfer for all
    settings is $q^2\,=\,-0.1\,GeV^2/c^2$.}
  \label{tab:kinematics}
  \center
  \begin{tabular*}{\columnwidth}{cccccc}
    \hline
    $\epsilon$ & $E_0$ & $E'$  & $p_d$       &$\theta_e$ & $\theta_d$\\
               & $[$MeV$]$ & $[$MeV$]$ & $[$MeV/c$]$ &           &\\
    \hline
    0.854 & 720 & 554 & 339 & $29.00^\circ$ & $48.69^\circ$\\
    0.590 & 435 & 269 & 339 & $55.10^\circ$ & $38.06^\circ$\\
    0.364 & 345 & 178 & 339 & $79.22^\circ$ & $29.40^\circ$\\ 
    \hline
  \end{tabular*}
\end{table}

A high power liquid deuteron target was used at luminosities of
$15\,\mathrm{MHz/\mu b}$, limited by the current in the drift chambers of the
deuteron spectrometer. Special care had to be taken to minimize the pathlength
of the deuterons in the target material. A long narrow cell of 4.8\,cm length
and 1\,cm width was shifted towards the side of the scattered electrons. In
this way an average pathlength of 3\,mm for the deuterons in the target
material could be achieved. In addition the beam had to be moved by a fast
magnet across the target area in a time scale of several kHz to avoid boiling
of the liquid deuterium. The wall of the target was built of a $10\,\mu$m
Havar foil.

To further reduce multiple scattering and energy loss the vacuum of the
scattering chamber was connected with the vacuum of the deuteron
spectrometer. For the electron detection a focal plane detector consisting of
4 layers of vertical drift chambers for spatial resolution and two layers of
scintillators for coincidence trigger and time of flight measurement was
used. A halocarbon gas \v{C}erenkov detector with an efficency of 99.8\% was
used for the separation of electrons and charged pions dominantly produced off
the target walls. For the deuteron detection, only one layer of scintillators
after four layers of vertical drift chambers could be used, since the
deuterons were already stopped after a pathlength of a few millimeters in the
first scintillator layer.

The large energy loss and the deuteron loss due to nuclear reactions of the
low energy deuterons made it necessary to monitor the efficiency of the
deuteron detection very carefully. In order to estimate these effects we used
the standard formulas for the energy loss \cite{leo} and the calculations of
\cite{pitz} for the contributions of nuclear reactions. Since these
calculations are only valid with large restrictions in our energy range, we
checked and calibrated in addition the deuteron detection efficiency by a
coincidence measurement of the scattered electron and the deuteron in the
elastic d(e,e'd) reaction. We used several settings of the elastic line at
different positions of the focal plane and compared the results with the known
cross section in the parameterization of \cite{platch}.

\begin{figure}
  \center \resizebox{7.5cm}{!}{\includegraphics{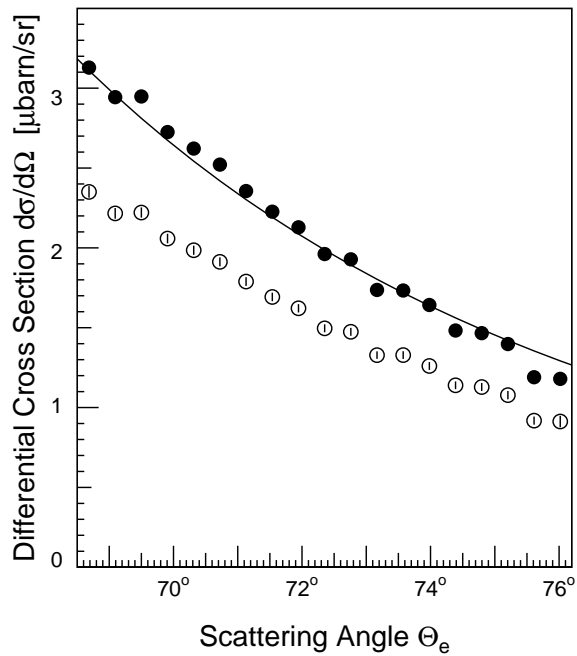}}
  \caption{
    The elastic cross section of the d(e,e'd) reaction at an incident energy
    of $E_{0}=420\,\mathrm{MeV}$. The solid line shows the known cross section
    calculated with the parameterization of \cite{platch}. The open circles
    show the coincidence measurement without efficiency correction, the solid
    circles show the measured cross section after full efficiency correction.}
  \label{fig:elastic}
\end{figure}

As expected from extensive simulations on the computer, we had to apply
correction factors of the order of up to 1.5 for the lowest deuteron energies.
Figure~\ref{fig:elastic} shows this correction for the elastic
measurements in comparison with the calculated elastic cross section.

\subsection{Reaction Identification}
The reaction $d(e,e'd)\pi^0$ was identified in two steps. First, the
coincidence time was determined by measuring the time of flight of the
deuteron and the electron and correcting it for the reconstructed path length
inside the spectrometers and the measured momenta. The coincidence time
resolution of 3.3\,ns FWHM was limited by the uncertainty in the flight path
reconstruction because of the large multiple scattering of the low energy
deuterons.

\begin{figure}
  \center \resizebox{7.5cm}{!}{\includegraphics{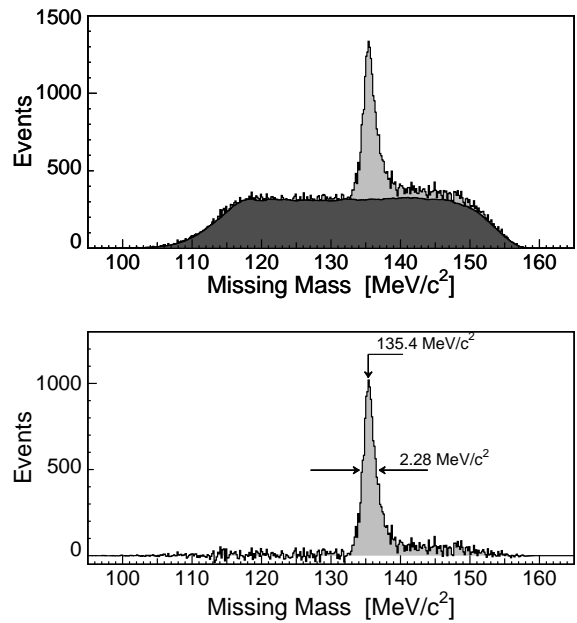}}
  \caption{The missing mass distribution with a cut on the coincidence time
    (total) and with a cut on a timing window shifted by 5\,ns besides the
    coincidence time peak for the background (dark gray). The lower picture
    shows the background subtracted missing mass spectrum. The radiative tail
    appears at higher masses.}
  \label{fig:missingmass}
\end{figure}
After the timing cut, the missing mass was calculated from the measured
four-vectors of incident electron $e$, scattered electron $e'$, and initial and
final deuteron $d$ and $d'$ by $m_{\mathrm{miss}}^2=(e+d-e'-d')^2$.
Figure~\ref{fig:missingmass} shows the distribution of the missing mass before
and after background subtraction. As can be seen, the $\pi^0$ can be
identified with a missing mass resolution of $2.28\,\mathrm{MeV/c^2}$ FWHM,
which is again determined by the multiple scattering of the low energy
deuterons in the wire chambers.

The effect of the radiative tail at higher missing masses was corrected by a
simulation using the standard formulas of \cite{motsai} in the peaking
approximation.

Since the recoil deuteron is measured directly, no model dependent correction
for the deuteron break up is necessary.

\subsection{Systematic errors}
The error of this experiment is dominated by the systematic errors. 
As known for threshold measurements the calibration of the measured
electron momentum which is almost proportional to the center-of-mass energy
causes the largest contribution.

The systematic error caused by the efficiency correction of the deuteron
detection could be checked by our elastic calibration measurements and was
determined to be 1.7\%. Further the detector efficiencies, contributions of
cuts and phase space integration, errors of the luminosity summation, and
condensation on the cold target walls were taken into account.

The total systematical errors, together with the statistical errors, are
compiled in table~\ref{tab:errors}.
\begin{table}
  \caption{Statistical and systematical errors of the total cross
    section}
  \label{tab:errors}
  \begin{tabular*}{\columnwidth}{crrrrrr}
    \hline
    Setting 
    & \multicolumn{2}{c}{$\epsilon=0.854$} 
    & \multicolumn{2}{c}{$\epsilon=0.590$}
    & \multicolumn{2}{c}{$\epsilon=0.364$}\\
    \hline 
    $\Delta W$ &stat.&sys.&stat.&sys.&stat.&sys.\\
    $[$MeV$]$& $[\%]$ & $[\%]$ & $[\%]$ & $[\%]$ & $[\%]$ & $[\%]$ \\
    \hline
    $0.5$ & 7.5 & 22.7 & 6.9 & 20.3 & 7.1 & 22.9 \\
    $1.5$ & 3.7 & 12.7 & 3.2 &  9.1 & 3.8 & 10.4 \\
    $2.5$ & 2.9 &  6.0 & 2.8 &  3.3 & 3.0 &  4.1 \\
    $3.5$ & 2.8 &  3.6 & 3.0 &  3.7 & 3.2 &  4.8 \\
    \hline
  \end{tabular*}
\end{table}

\section{Discussion of the Results}

\begin{figure}
  \center \resizebox{7.5cm}{!}{\includegraphics{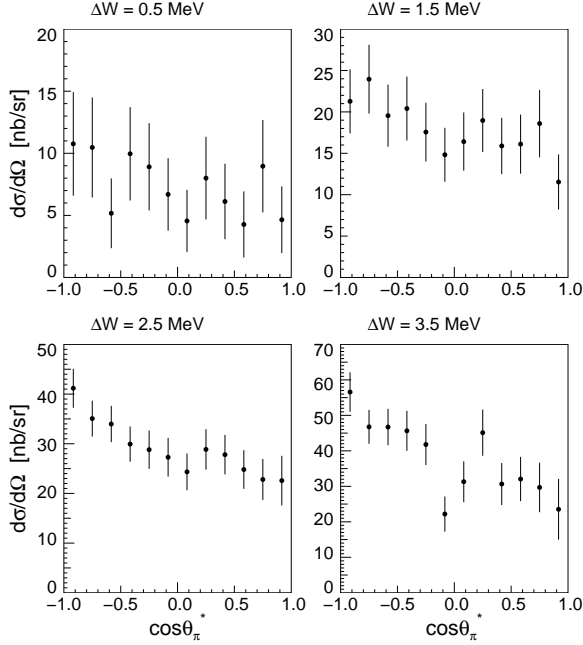}}
  \caption{The differential cross section for $\epsilon = 0.590$.}
  \label{fig:wq22}
\end{figure}

\begin{figure}
  \center \resizebox{7.5cm}{!}{\includegraphics{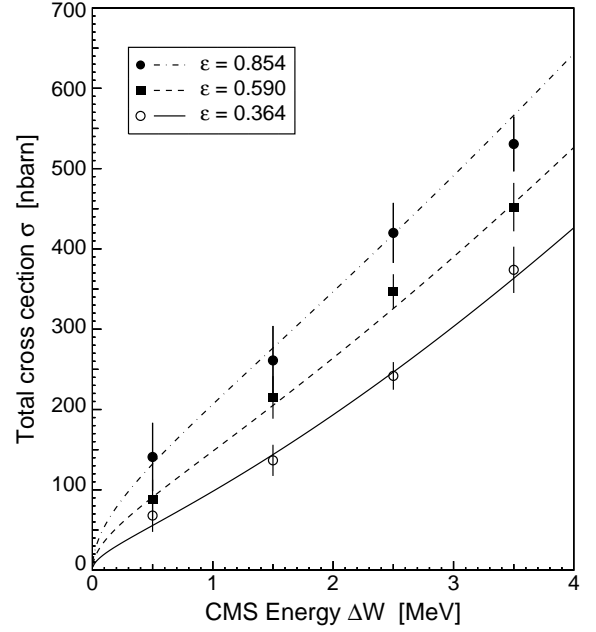}}
  \caption{The total cross section for three different values of the photon
    polarization $\epsilon$. The lines show the result of a least squares
    fit with the assumption of only $p$ and $s$ waves contributing to the
    cross section near threshold.  }
  \label{fig:total}
\end{figure}

The measured differential cross section is compiled in
table~\ref{tab:data}. For one setting ($\epsilon=0.590$),
figure~\ref{fig:wq22} shows the data points, including the combined
statistical and systematical errors. As expected, only angular structures up
to $\mathcal{O}(\cos^2\theta_\pi^*)$ appear and justify a fit with the
assumption of only $s$ and $p$ waves contributing to the cross
section. Figure~\ref{fig:total} shows the total cross section and the result
of a $\chi^2$ fit to the data with the assumption of a constant $s$ wave
amplitude and $p$ wave amplitudes rising linearly\footnote{In this energy
range the difference between a linear behavior as found empirically
\cite{Berg98} and the theoretically predicted behavior $\sim p_\pi^*/\omega$
can be neglected.}  with $p_\pi^*$. At the precision of these data, no cusp
effects of the opening deuteron break up threshold at 2.2\,MeV above threshold
can be observed.

At present, only threshold calculations are available in the framework of
ChPTh \cite{BKrM99} which provide a prediction for the $s$ wave multipoles. 
From the fit we can extract the $s$ wave multipoles at threshold through the
reduced threshold $s$ wave cross section
\begin{eqnarray}
a_0 &=& 
\left. \frac38\frac{k_\gamma^*}{p_\pi^*}\frac{d\sigma}{d\Omega_\pi^*}
 \right|_{\Delta W \rightarrow 0}
  = |\Ed|^2 + \epsilon_\mathrm{L}|\Ld|^2
\end{eqnarray}
Since $\epsilon_\mathrm{L}\approx 9\epsilon$ the longitudinal amplitude \Ld\ 
contributes with a much higher weight to the $s$ wave cross section
than \Ed{} and only an upper limit of one standard
deviation can be extracted for \Ed{}. From this analysis one gets
\begin{eqnarray*}
  |\Ed| &\le&  0.42           \cdot 10^{-3}/m_\pi\\ 
  |\Ld| & = & (0.50 \pm 0.11) \cdot 10^{-3}/m_\pi
\end{eqnarray*}

\begin{figure}
  \center
  \resizebox{7.5cm}{!}{\includegraphics{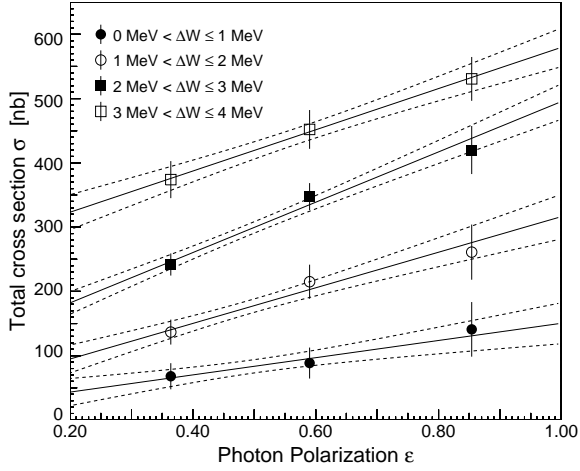}}
  \caption{Rosenbluth plot with straight line fits including one standard
  deviation error band.}
  \label{fig:rosenbluth}
\end{figure}

The classical Rosenbluth method, i.e. the total cross section plotted against
the photon polarization $\epsilon$, represents a consistency check
(Figure~\ref{fig:rosenbluth}). For each energy bin a straight line fit was
performed to separate $f_\mathrm{T}$ by its offset and $f_\mathrm{L}$ by its
slope. At threshold, only $|\Ld|^2$ contributes to the longitudinal cross
section and can be determined by an extrapolation of $f_\mathrm{L}$ to the
threshold point. By this technique $|\Ld|=(0.47\pm 0.18) \cdot
10^{-3}/m_{\pi}$ is extracted in good agreement with the previous
analysis. Again, the kinematically suppressed multipole \Ed\ cannot be
determined.

\begin{figure}
  \center \resizebox{7.5cm}{!}{\includegraphics{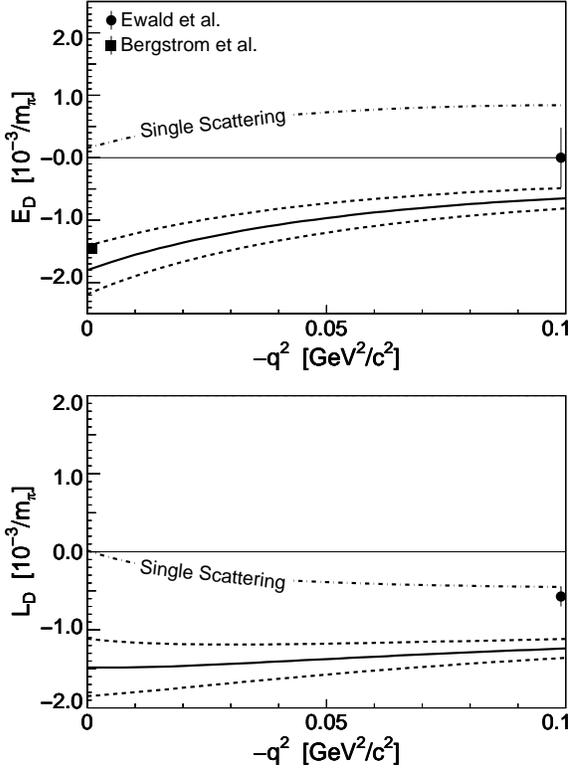}}
  \caption{
    The extracted $s$ wave multipoles (circles) in comparison with the
    prediction of Chiral Perturbation Theory \cite{BKrM99}.  The photon
    point of \cite{Berg98} is plotted as a square. For the explanation of the
    curves see text.}
  \label{fig:chptcomparison}
\end{figure}

Figure~\ref{fig:chptcomparison} shows the extracted $s$ wave multipoles in
comparison with the prediction of Chiral Perturbation Theory \cite{BKrM99}.
These calculations are performed to third order in the chiral expansion and
are shifted to reproduce the result of the fourth order calculation at the
photon point \cite{BBLM97}.  The solid line shows the full calculation, for
the dashed lines the calculated free neutron amplitude was varied by
$\pm 10^{-3}/m_\pi$ to indicate the sensitivity of the calculation to this
amplitude. The dash-dotted line shows the calculation without two body
currents, i.e. only the amplitudes of the free nucleons folded with the
pertinent deuteron form factors are included and no pion exchange between the
two nucleons was taken into account. For this picture, we assumed the sign for
\Ld\ to be the same as calculated in ChPTh.

As stated in \cite{Berg98}, their transverse multipole \Ed\ is already at the
photon point $\Delta\Ed = 0.35\,\cdot 10^{-3}/m_\pi$ above the ChPTh
prediction \cite{BBLM97}. A shift by this amount would make our result
for \Ed\ consistent with the calculations of \cite{BKrM99}.

\section{Summary}

A first measurement of the coherent threshold electroproduction of neutral
pions off the deuteron was performed and the differential cross section could
be determined up to 4\,MeV above threshold for three different values of the
photon polarization. The absolute values of the $s$ wave multipoles were
extracted and compared to the predictions of ChPTh. Although smaller than
expected, our extracted upper limit for $|\Ed|$ is consistent with ChPTh
calculations.  $|\Ld|$ is overestimated by the theory by a factor of 2. More
dramatic is this discrepancy in terms of cross sections: The reduced threshold
$s$ wave cross section $a_0$ is one order of magnitude smaller than expected.

Both, in photo- and in electroproduction the measured differential cross
sections allow tests of predictions for the contributing $p$ waves which are
not yet calculated in the framework of ChPTh.

\begin{table}
  \caption{The differential cross section. The cross section is integrated
    over the complete accepted out-of-plane angle $\phi_\pi^*$.}
  \label{tab:data}
  \center
  \def\arraystretch{1}
  \begin{tabular*}{\columnwidth}
    {rr@{\,$\pm$}rr@{\,$\pm$\,}rr@{\,$\pm$\,}rr@{\,$\pm$\,}r}
    \hline
    & \multicolumn{6}{c}{$d\sigma/d\Omega^*_\pi$~~~~~[nb/sr]} \\
    $\cos \theta_\pi^*$ 
    & \multicolumn{2}{c}{$\epsilon=0.854$}
    & \multicolumn{2}{c}{$\epsilon=0.590$}
    & \multicolumn{2}{c}{$\epsilon=0.364$}\\
    \hline
    \multicolumn{7}{c}{$0\,\mathrm{MeV}< \Delta W \le1\,\mathrm{MeV}$}\\
    $-0.917$ & 16.79 &  7.65& 10.76 &  4.17&  3.94 &  2.52\\
    $-0.750$ & 20.14 &  8.22& 10.47 &  4.03&  4.37 &  2.58\\
    $-0.583$ & 21.73 &  8.36&  5.17 &  2.81&  2.48 &  2.00\\
    $-0.417$ &  6.45 &  4.46&  9.96 &  3.76&  8.07 &  3.30\\
    $-0.250$ & 10.54 &  5.19&  8.91 &  3.51&  6.25 &  2.78\\
    $-0.083$ & 15.04 &  6.42&  6.69 &  2.92&  7.38 &  3.00\\
    $ 0.083$ &  6.22 &  3.98&  4.55 &  2.51&  4.52 &  2.28\\
    $ 0.250$ &  9.97 &  4.98&  8.00 &  3.33&  8.47 &  3.35\\
    $ 0.417$ &  7.40 &  4.50&  6.12 &  3.04&  5.00 &  2.45\\
    $ 0.583$ & 13.89 &  5.95&  4.27 &  2.66&  8.08 &  3.31\\
    $ 0.750$ &  8.43 &  4.66&  8.96 &  3.72&  5.16 &  2.51\\
    $ 0.917$ &  4.31 &  3.18&  4.65 &  2.69&  4.24 &  2.02\\
    \hline
    \multicolumn{7}{c}{$1\,\mathrm{MeV}< \Delta W \le2\,\mathrm{MeV}$}\\
    $-0.917$ & 28.14 &  6.43& 21.28 &  3.87& 16.13 &  3.31\\
    $-0.750$ & 24.94 &  5.96& 23.95 &  4.16& 17.51 &  3.46\\
    $-0.583$ & 29.05 &  6.55& 19.55 &  3.76& 11.71 &  2.69\\
    $-0.417$ & 22.11 &  5.50& 20.41 &  3.86& 11.97 &  2.70\\
    $-0.250$ & 30.73 &  6.75& 17.56 &  3.55& 10.05 &  2.45\\
    $-0.083$ & 25.77 &  6.12& 14.82 &  3.26& 11.29 &  2.69\\
    $ 0.083$ & 16.52 &  4.85& 16.41 &  3.52& 11.60 &  2.71\\
    $ 0.250$ & 15.79 &  4.73& 18.96 &  3.80& 12.74 &  2.90\\
    $ 0.417$ & 17.36 &  5.18& 15.89 &  3.40& 10.54 &  2.59\\
    $ 0.583$ & 15.60 &  4.92& 16.11 &  3.57& 10.16 &  2.62\\
    $ 0.750$ & 15.55 &  5.17& 18.59 &  4.07&  7.38 &  2.27\\
    $ 0.917$ & 19.54 &  5.97& 11.53 &  3.33&  5.59 &  1.99\\
    \hline
    \multicolumn{7}{c}{$2\,\mathrm{MeV}< \Delta W \le3\,\mathrm{MeV}$}\\
    $-0.917$ & 50.15 &  6.42& 41.17 &  3.93& 27.56 &  3.33\\
    $-0.750$ & 40.63 &  5.67& 35.07 &  3.60& 26.64 &  3.22\\
    $-0.583$ & 44.79 &  6.01& 33.97 &  3.66& 24.83 &  3.04\\
    $-0.417$ & 37.20 &  5.55& 29.93 &  3.55& 22.36 &  2.90\\
    $-0.250$ & 43.22 &  6.31& 28.80 &  3.85& 24.67 &  3.13\\
    $-0.083$ & 36.12 &  5.79& 27.28 &  3.90& 17.91 &  2.90\\
    $ 0.083$ & 36.45 &  5.96& 24.34 &  3.70& 17.80 &  2.98\\
    $ 0.250$ & 28.86 &  5.46& 28.86 &  4.07& 22.75 &  3.33\\
    $ 0.417$ & 31.57 &  5.97& 27.80 &  3.97& 17.19 &  2.94\\
    $ 0.583$ & 31.43 &  6.61& 24.81 &  3.89& 13.91 &  2.72\\
    $ 0.750$ & 19.19 &  5.80& 22.81 &  4.13& 15.64 &  2.89\\
    $ 0.917$ & 20.34 &  7.45& 22.58 &  4.99& 10.57 &  2.66\\
    \hline
    \multicolumn{7}{c}{$3\,\mathrm{MeV}< \Delta W \le4\,\mathrm{MeV}$}\\
    $-0.917$ & 70.78 &  6.66& 56.58 &  5.57& 49.62 &  5.51\\
    $-0.750$ & 63.13 &  6.18& 46.77 &  4.78& 36.83 &  4.41\\
    $-0.583$ & 56.53 &  6.06& 46.71 &  5.12& 33.87 &  4.33\\
    $-0.417$ & 47.84 &  6.02& 45.64 &  5.64& 35.88 &  4.82\\
    $-0.250$ & 57.50 &  7.04& 41.79 &  5.78& 34.65 &  5.08\\
    $-0.083$ & 55.06 &  7.23& 22.19 &  4.95& 22.52 &  4.40\\
    $ 0.083$ & 41.72 &  6.70& 31.30 &  5.77& 32.23 &  5.26\\
    $ 0.250$ & 41.41 &  7.06& 45.12 &  6.52& 25.87 &  4.89\\
    $ 0.417$ & 26.78 &  6.49& 30.65 &  5.92& 23.35 &  4.76\\
    $ 0.583$ & 24.05 &  6.84& 32.06 &  6.23& 27.31 &  4.74\\
    $ 0.750$ & 29.06 &  8.52& 29.70 &  6.98& 28.11 &  4.99\\
    $ 0.917$ & 16.78 & 10.93& 23.53 &  8.55& 23.70 &  5.06\\
\hline
\end{tabular*}
\end{table}

\ack

This work was possible thanks to the excellent performance of the Mainz
Microtron MAMI. It was supported by the special research project SFB~443 of
the Deutsche Forschungsgemeinschaft (DFG) and the Federal State of
Rhineland-Palatinate. A.\,M.~Bernstein is grateful to the Alexander von
Humboldt Foundation for a Humboldt Research Award.


\begin{thebibliography}{99}
\def\Journal#1#2#3#4{#1 {\bf #2}, (#3) #4}
\def\EPJ#1#2#3{\Journal{Eur. Phys. J.}{#1}{#2}{#3}}
\def\PRL#1#2#3{\Journal{Phys. Rev. Lett.}{#1}{#2}{#3}}
\def\PR#1#2#3{\Journal{Phys. Rev.}{#1}{#2}{#3}}
\def\PRC#1#2#3{\Journal{Phys. Rev. C}{#1}{#2}{#3}}
\def\PL#1#2#3{\Journal{Phys. Lett.}{#1}{#2}{#3}}
\def\NP#1#2#3{\Journal{Nucl. Phys.}{#1}{#2}{#3}}
\def\NPA#1#2#3{\Journal{Nucl. Phys. A}{#1}{#2}{#3}}
\def\NIM#1#2#3{\Journal{Nucl. Instr. and Meth.}{#1}{#2}{#3}}
\bibitem{chpt94}   A.\,M.~Bernstein, B.\,R.~Holstein (Eds.), 
                   Lect. Notes in Phys. 452 (1995)
\bibitem{chpt97}   A.\,M.~Bernstein, D.~Drechsel, Th.~Walcher (Eds.),
                   Lect. Notes in Phys. 513 (1998)
\bibitem{nikhef}   H.\,B.~van~den~Brink et al., \PRL{74}{1995}{3561}
\bibitem{distler}  M.\,O.~Distler et al.,   \PRL{80}{1998}{2294}
\bibitem{BKMe96}   V.~Bernard, N.~Kaiser, U.-G.~Mei\ss{}ner,
                   \NPA{607}{1996}{379-401};
                   Erratum \NPA{633}{1998}{695-697}
\bibitem{Berg98}   J.~C.~Bergstrom et al., \PRC{57 6}{1998}{3203}
\bibitem{BBLM97}   S.\,R.~Beane et al., \NPA{618}{1997}{381}
\bibitem{eberts}   Th.~Ebertsh\"auser, H.~Arenh\"ovel, 
                   \EPJ{A6}{1999}{431-443}
\bibitem{aren}     H.~Arenh\"ovel, Few Body Syst. 27 (1999) 141-162
\bibitem{dreispek} K.\,I.~Blomqvist et al., \NIM{A403}{1998}{263}
\bibitem{leo}      W.\,R. Leo, Techniques for Nuclear and Particle Physics
  Experiments, Springer 1987, and C.~Caso et al., \EPJ{C3}{1998}{1-794}
%\bibitem{hencken}  K.~Hencken, private communication
\bibitem{pitz}     D.~Pitz, PhD thesis, CEN Saclay, Gif sur Yvette 1999
\bibitem{platch}   S.~Platchkov et al., \NPA{510}{1990}740
\bibitem{motsai}   L.\,W.~Mo, Y.\,S.~Tsai, Rev. Mod. Phys 41 (1969) 205
\bibitem{BKrM99}   V.~Bernard, H.~Krebs, U.-G.~Mei\ss{}ner, 
                   \PRC{61}{2000}{58201}
\end{thebibliography}
\end{document}